\documentclass[pre,superscriptaddress,reprint]{revtex4-1}

\usepackage{amsmath}
\usepackage{graphicx}
\usepackage{bm}
\usepackage{MnSymbol}
\usepackage{eufrak}
\usepackage{amsfonts} 
\usepackage{float}
\usepackage[usenames,dvipsnames,x11names]{xcolor}
\usepackage[normalem]{ulem}
\usepackage{tikz}
\usetikzlibrary{arrows}
\usepackage{mathrsfs}

\newcommand{\EValTot}{w}

\begin{document}

\title{Synchronized states in dissipatively coupled harmonic oscillator networks}

\author{Juan N.\ Moreno}
\email{moreno@pks.mpg.de}
\affiliation{Max Planck Institut für Physik komplexer Systeme, Nöthnitzer Str.\ 38, 01187 Dresden, Germany}
\author{Christopher W.\ W\"achtler}
\email{cwwaechtler@berkeley.edu}
\affiliation{Max Planck Institut für Physik komplexer Systeme, Nöthnitzer Str.\ 38, 01187 Dresden, Germany}
\affiliation{Department of Physics, University of California, Berkeley, California 94720, USA}
\author{Alexander Eisfeld}
\email{eisfeld@pks.mpg.des}
\affiliation{Max Planck Institut für Physik komplexer Systeme, Nöthnitzer Str.\ 38, 01187 Dresden, Germany}
\affiliation{Universität Potsdam, Institut für Physik und Astronomie,
Karl-Liebknecht-Str.\ 24-25, 14476 Potsdam, Deutschland}

\begin{abstract}
The question under which conditions oscillators with slightly different frequencies synchronize appears in various settings. 
We show that synchronization can be achieved even for \textit{harmonic} oscillators that  are bilinearly coupled via a purely dissipative interaction.
By appropriately tuned gain/loss   stable dynamics may be achieved where for the cases studied in this work all oscillators are synchronized.
These findings are interpreted using the complex eigenvalues and eigenvectors of the non-Hermitian matrix describing the dynamics of the system. 
\end{abstract}

\maketitle

\section{Introduction}
Synchronization is a fascinating phenomenon, which can be interpreted as a display of cooperative behavior appearing in many complex systems \cite{pikovsky2003synchronization, StrogatzBook2018}.
Since the first observation by Huygens in the late 1600s \cite{bennett2002huygens}, it has been studied in diverse communities, where it plays an important role in our understanding for example in electric networks in engineering, circadian rhythms in biology, pattern formation in statistical mechanics, and chemical reactions in chemistry \cite{Strogatz1993, Rosenblum2003, arenas2008synchronization}. By now, it is seen as a universal phenomenon that is important both in fundamental studies and in technical applications, ranging from laser networks \cite{thornburg1997chaos}, to phase-locked loops \cite{lynch1995mode}, Josephson junction arrays \cite{cawthorne1999synchronized,fazio2001quantum}, spin-torque resonators \cite{slavin2009spin}, and power grids \cite{Nishikawa_2015}. Even today, the originally observed phenomenon of clock synchronization remains a crucial application for modern communication networks \cite{bellamy1995digital,narula2018requirements}. 
 
Typically synchronization is viewed in terms of the adjustment of rhythms of autonomous oscillators, which attain stable periodic orbits without active regulation from the outside \cite{JenkinsPR2013} and thus require nonlinearities in the governing equations of motion. 
Far less common is the investigation of synchronization in  models that are linear in both the oscillators and the couplings. 
 Without dissipation, coupled harmonic oscillators form collective eigenmodes, where the individual oscillators perform motion with a fixed phase relation. However, a system not initialized in an eigenmode usually stays in a superposition of several eigenmodes with different eigenfrequencies resulting in a beating pattern. Moreover, if the number of coupled oscillators is large, the system dynamics does not need to exhibit perfect revivals in general and synchronized motion is absent. 
Hence in a closed system of oscillators, only for an eigenmode as initial condition one obtains a time-independent phase relation between the oscillators.  
However if the system is not closed, but subject to gain and loss, the open system dynamics allow for a situation where all eigenmodes but one are damped. 
Then, synchronization is possible as long as the respective eigenstate is present in the initial state. 
However, in order to achieve a situation where all but one mode are damped, one needs to carefully balance gain and loss. 

In contrast to a self-sustained system where the nonlinearity counteracts the dissipation (or gain) in order to stabilize periodic orbits, a {\it single} linear harmonic oscillator only exhibits the following dynamics in the absence of periodic driving: Either the dissipation exceeds the gain, such that the amplitude of the dissipative systems shrinks and eventually reaches a single point in phase space, or in other way around, where the gain exceeds the dissipation, the oscillation amplitude infinitely grows. In the special case where both are equivalent the system is effectively described by closed system dynamics with infinitely many closed orbits in phase space depending on the initial energy of system. However, when coupling between linear oscillators are introduced, many more solutions are possible. 

Here, we investigate a network of linear harmonic oscillators subject to gain and loss. Generally, one would consider each oscillator to couple to its own environment and direct coupling between two or more entities in the network. However, a purely dissipative coupling leads to intriguing phenomena also for self-sustained oscillators like for example oscillator death \cite{pikovsky2003synchronization}. In our model of linear oscillators, it allows for the emergence of dissipation free subspaces in parameter space. Within these subspaces we find  periodic motion of all oscillators in the network, that is starting from an (nearly) arbitrary initial state the system reaches a regime during time propagation in which all oscillators exhibit synchronized motion for a long time. At this point, let us specify the notion of synchronization we use throughout this work:
\\
--- With 'long time' we mean times long compared to the  eigenfrequencies of the individual oscillators and we focus on the case where all oscillators have small deviations from a common 'mean frequency'. 
In the ideal case they oscillate forever.
    \\
    --- With 'synchronoized' we mean that the oscillators have a fixed phase relation. 
    Ideally we want that all oscillators have the same amplitude. 
    If this is the case, then we denote it as {\it full synchronization}. 
    If the system is not in a fully synchronized state, we will characterize its {\it degree of synchronization} by a suitable measure.
    \\
    --- With 'arbitrary' initial state we mean that for most initial states synchronization is achieved, yet there exist 
   some special initial conditions that do not lead to synchronization. 

We note that within the above definitions for \text{uncoupled} oscillators one only finds synchronization, when there is no gain and loss and all oscillators have the same frequency.

The remainder of the paper is organized as follows: In Sec.~\ref{subsec:GeneralConsiderations} we summarize some general considerations of synchronization for linearly coupled harmonic oscillators important for our work, followed by the specific model under investigation in Sec.~\ref{subsec:OurModel}. In the subsequent Sec.~\ref{sec:results} we discuss our results, which includes the special case of two coupled oscillators in Sec.~\ref{subsec:TwoOscillators} and the more general case of many oscillators in Sec.~\ref{subsec:NOscillators}. Finally, we conclude in Sec.~\ref{sec:Conclusions}.

\section{Model and basic formalism}
\subsection{General considerations of synchronization in linear oscillator models}
\label{subsec:GeneralConsiderations}
To introduce the basic concepts and notation, we consider $N$ harmonic oscillators in a network, each labeled by a subscript $n=1,...,N$.
The motional state of each oscillator is  characterized by a time dependent complex amplitude $a_n(t)=|a_n(t)|\mathrm{e}^{\mathrm{i}\phi_n(t)} $. 
If all oscillators in the network oscillate with a common  real frequency $\omega_\mathrm{syn}$ while their relative amplitudes remain constant, we will refer to it as synchronization. Using a vector notation $\vec{a}(t)= \left[a_1(t), ..., a_N(t)\right]^\intercal$, such synchronized motion may be  expressed as
\begin{equation}
\label{eq:IdealSync}
   \vec{a}(t)={f(t)} \vec{a}_\mathrm{syn}  \mathrm e^{-\mathrm i \omega_\mathrm{syn} t},
\end{equation}
where $f(t)$ is a real function that takes into account the possibility that the amplitudes decay (or grow) over time, which we will discuss in Sec.~\ref{subsec:OurModel} in more detail.
In the case  of $f(t)=1$ the motion represents a periodic steady state, which we refer to  as \textit{ideal synchronized motion}. 

\begin{figure}
    \centering
    \includegraphics[width=0.8\columnwidth]{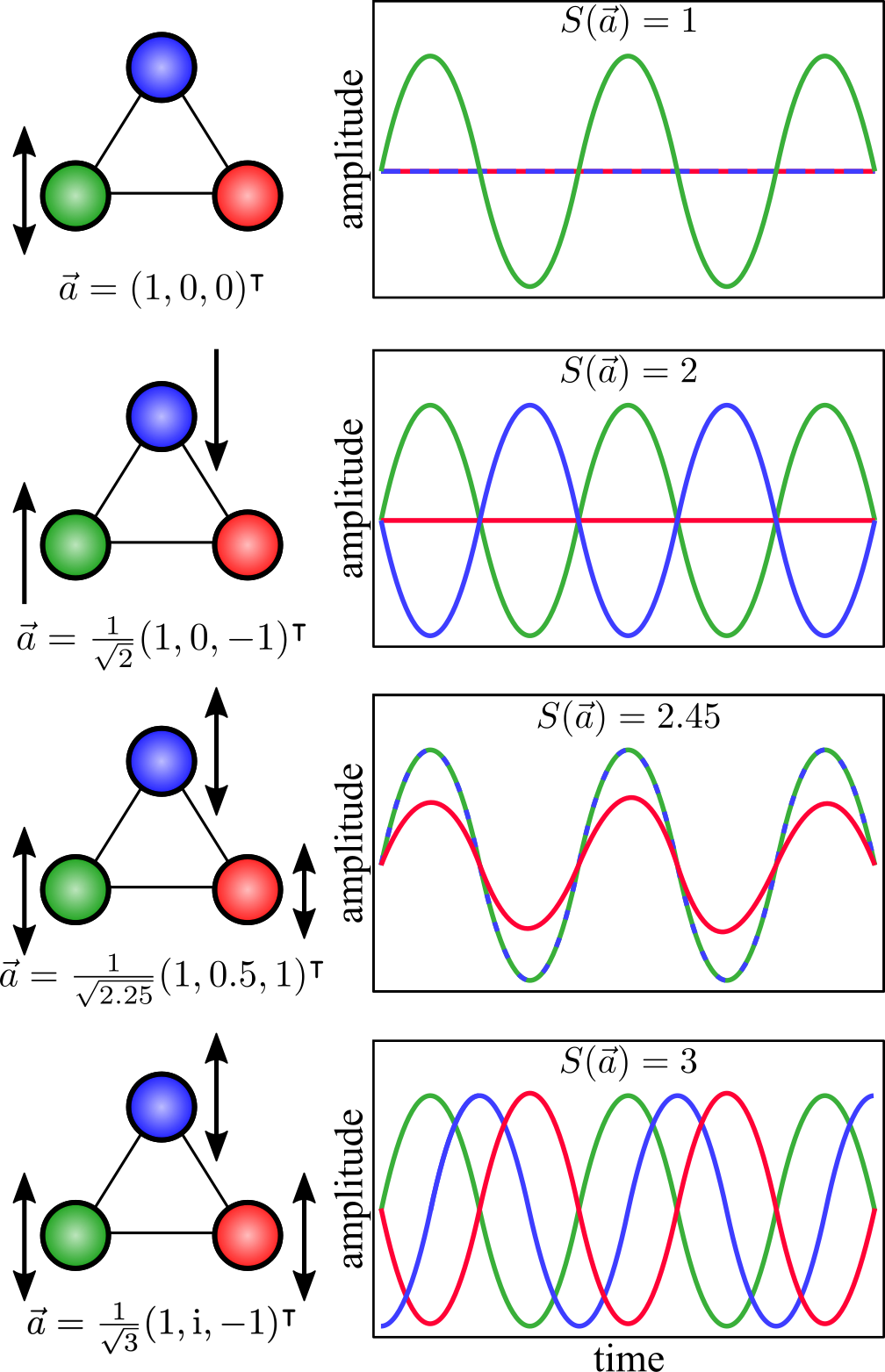}
   \caption{Illustration of potentially attainable synchronized motion in a network of $N=3$ oscillators. The inverse participation ratio $S(\vec{a})$ increases from top to bottom in accordance with the transition from partially to fully synchronized motion.} 
    \label{fig:Example}
\end{figure}

The above notion is not sufficient to fully characterize synchronized motion as for example a single oscillatory site in the network (while all other oscillators are at rest) also fulfills Eq.~(\ref{eq:IdealSync}). 
It is thus necessary to also quantify the \textit{degree of synchronization} of a vector $\vec{a}$, which we denote by $\mathcal{S}(\vec{a})$. 
To this end, we use the inverse participation ratio \cite{kramer1993localization}
\begin{equation}
\label{eq:IPR}
  \mathcal{S}(\vec{a})=\frac{1}{\sum_{n=1}^N|a_{n}|^4},
\end{equation}
which takes values between $1$ and $N$. Here, a value of $\mathcal S = 1$ corresponds to the aforementioned case of a single oscillator in motion, whereas a value of $\mathcal{S}=N$ indicates \textit{fully synchronized motion}, i.e. all nodes have the same amplitude (without phase). 
Values of $\mathcal{S}=\tilde{N}<N$ correspond to \textit{partial synchronization} of approximately $\tilde{N}$ oscillators. 
In Fig.~\ref{fig:Example}, we illustrate different degrees of synchronization and their respective dynamics in a network of three oscillators.

The time evolution of a linearly coupled network of harmonic oscillators in the presence of gain and loss is generally expressed as 
\begin{equation}
\label{eq:DifferentialVecA}
    \frac{d}{dt} \vec{a}= -\mathrm i W \vec{a}, 
\end{equation}
where we assume the non-Hermitian matrix $W$ to be time-independent. Then, the state of the system at time $t$ is simply given by 
\begin{equation}
\label{eq:TimeEvolutionVecA}
    \vec{a}(t)= e^{-\mathrm i W t} \vec{a}(0),
\end{equation}
where $\vec{a}(0)$ denotes the initial state at time $t=0$. Thus, the dynamics of the network is fully characterized by the matrix $W$, in particular by its eigenvalues and eigenvectors. 
Since $W$ is (in general) non-Hermitian, there exist right and left eigenvectors defined via 
\begin{align}
    W \vec{c}_j=& w_j  \vec{c}_j\quad \text{and}\quad    \vec{z}_j^\dagger W = \vec{z}_j^\dagger w_j. 
\end{align}
Here, $\dagger$ indicates the complex conjugated and transpose, and the eigenvectors are normalized according to 
\begin{equation}
\begin{aligned}
    \vec{c}_j^\dagger \vec{c}_j &= 1 \quad \text{and}\quad \vec{z}_{j'}^\dagger \vec{c}_j = \delta_{j'j}.
\end{aligned}
\end{equation}
Note, that in general $\vec{c}_j^\dagger \neq \vec{z}_j^\dagger $. 
The matrix $W$ can now be  expressed  as $W=\sum_j \EValTot_j \vec{c}_j \vec{z}_j^\dagger$, such that the time evolution of Eq.~(\ref{eq:TimeEvolutionVecA}) is conveniently given by
\begin{align}
\label{eq:time-evol_eigen}
    \vec{a}(t)= \sum_j \vec{c}_j e^{-\mathrm i w_j t} \, \vec{z}_j^\dagger \vec{a}(0),
\end{align}
where $\vec{z}_j^\dagger \vec{a}(0)$ is the initial weight of the eigenstate $j$. While the real part of the complex eigenvalue $\EValTot_j$ determines the oscillation frequency of eigenmode $j$, the imaginary part  $\mathrm{Im}[\EValTot_j]$ determines  whether the oscillatory motion is damped ($\mathrm{Im}[\EValTot_j]<0$), growing ($\mathrm{Im}[\EValTot_j]>0$) or oscillates forever ($\mathrm{Im}[\EValTot_j]=0$). 

In order to obtain a time evolution of the form of Eq.~(\ref{eq:IdealSync}) with $f(t)=1$ after some initial transient time, i.e. dynamically reach the eigenstate with $\mathrm{Im}[\EValTot_\mathrm{sync}]=0$, the initial state needs to have non-vanishing overlap with the synchronized eigenstate [$\vec{z}_\mathrm{sync}^\dagger a(0)\ne 0$]. Furthermore, all other eigenstates present in the initial state need to have $\mathrm{Im}[\EValTot_j]<0$, such that they are damped.
In the following, we will therefore search for conditions and parameters under which {\it one} eigenstate fulfills $\mathrm{Im}[\EValTot_\mathrm{sync}]=0$ while all other eigenstates fulfill $\mathrm{Im}[\EValTot_j]<0$. Subsequently, we will characterize the degree of synchronization of the resulting state in terms of $\mathcal{S}$; cf. Eq.~(\ref{eq:IPR}).

\subsection{Linear oscillators with purely dissipative coupling \label{subsec:OurModel} }
After the general considerations of the previous Sec.~\ref{subsec:GeneralConsiderations}, let us now specify the network of interest throughout the remainder of this work: The individual oscillators have frequencies $\Omega_n\in\mathbb R$ and are arranged on a ring. Each oscillator is subject to gain/loss mediated via the rate $\gamma\in\mathbb R$ and interacts with its two nearest neighbors via a purely dissipative coupling $v\in\mathbb R$. 
For simplicity we assume that the coupling and dissipation is equal for all oscillators;
we are interested in the possibility of synchronization when the frequency of each oscillator is different, which corresponds to the  notion of synchronization as an adjustment of rhythms due to the presence of interactions. The equation of motion of the $n$-th oscillator is then given by 
\begin{align}
\label{eq:dot_a_k}
    \frac{d}{dt}{a}_n =& (-\mathrm{i}\Omega_n- {\gamma}) a_n
    - v(a_{n+1}
   + a_{n-1}),
\end{align}
with $a_0\equiv a_N$ and $a_{N+1}\equiv a_1$ to fulfill periodic boundary conditions. Note that positive values of $\gamma$ represent loss whereas negative  values correspond to gain. 
To simplify notation we express all energies in units of $v$ and take $v$ to be positive (the case of negative $v$ will be discussed later), i.e. $\omega_n=\Omega_n/v$, $g=\gamma/v$ and $\tau=t v$. Furthermore, we parameterize the frequencies as $\omega_n=\bar\omega +\Delta_n$. Then, Eq.~(\ref{eq:dot_a_k}) becomes
\begin{align}
\label{eq:dot_a_k_normalizedDelta}
    \frac{d}{d\tau}{a}_n =& [-\mathrm{i}(\bar{\omega}+\Delta_n)- {g}]a_n
    - (a_{n+1}
   + a_{n-1}).
\end{align}
Our goal in the following is to determine the values of $g$ for a given set of frequency differences $\Delta_n$, such that the oscillators perform synchronized motion in the sense discussed in Sec.~\ref{subsec:GeneralConsiderations}.

As the term $(-\mathrm{i}\bar{\omega} -g)$ is independent of the oscillator index $n$, it only trivially contributes to the overall dynamics; specifically oscillations with frequency $\bar{\omega}$ and damping/growing with rate $g$.  In matrix representation, Eq.~(\ref{eq:dot_a_k_normalizedDelta}) can be written in the form of Eq.~(\ref{eq:DifferentialVecA}) with $t\to\tau$ and $W=  (\bar\omega - \mathrm{i} g)\mathbb{I} + M$, where
\begin{align}
\label{eq:M_matrix}
    M=
    \begin{pmatrix}
     \Delta_1 & -\mathrm{i} &0 &\dots & -\mathrm{i}\\
    -\mathrm{i} &  \Delta_2 & -\mathrm{i} &\dots &0
    \\
    0 & -\mathrm{i} 
    \\
    \vdots
    \\
    -\mathrm{i} &0 & \dots &-\mathrm{i}& \Delta_N 
    \end{pmatrix}
\end{align}
Note, that the (left and right) eigensvectors of $W$ and $M$ are identical and their eigenvalues are simply shifted, i.e., if $M\vec{c}_j = \lambda_j \vec{c}_j$ then $W\vec{c}_j=w_j \vec{c}_j$ with 
\begin{equation}
\label{eq:w_posv}
w_j = \bar{\omega} +  \mathrm{Re}[\lambda_j] +\mathrm{i} (-g+\mathrm{Im}[\lambda_j]),
\quad v>0.
\end{equation}
Moreover, as $M$ only depends on $\Delta_n$, the eigenvectors and thus the degree of synchronization $\mathcal{S}(\vec{c})$ is independent of $g$.

Let us summarize the general conditions of the previous Sec.~\ref{subsec:GeneralConsiderations} for synchronized motion tailored to the specifics of our system discussed here:
\begin{itemize}
\item[(i)] There exists a single eigenstate $\vec{c}_\mathrm{sync}$ of $W$ with purely real eigenvalue. This corresponds to a state $\vec{c}_\mathrm{sync}$ that fulfills $-g+\mathrm{Im}[\lambda_\mathrm{sync}] =0$, where $M\vec{c}_\mathrm{sync}=\lambda_\mathrm{sync}\vec{c}_\mathrm{sync}$.
\item[(ii)] All other eigenstates of $W$ have negative imaginary part for the set of parameters determined in (i). That corresponds to $-g+\mathrm{Im}[\lambda_j]<0$
for all $j \ne \mathrm{sync}$.
\item[(iii)] The synchronization measure $\mathcal{S}(\vec{c}_\mathrm{sync})$ should be as large as possible. 
Ideally $\mathcal{S}(\vec{c}_\mathrm{sync})=N$.
\end{itemize}

So far, we have taken $v$ to be positive.
For negative values of $v$ we define the scaled energies in terms of $-v$ such that $\omega_n=\bar\omega +\Delta_n=-\Omega_n/v $, $g=-\gamma_n/v$, and $\tau=-tv$. Then, Eq.~(\ref{eq:dot_a_k_normalizedDelta}) becomes 
\begin{equation} 
\frac{d}{d\tau}{a}_n = [-\mathrm{i}(\bar\omega +\Delta_n) - g]a_n
    + (a_{n+1}
   + a_{n-1}),
\end{equation}
where the first term remains identical while the sign changes in front of the oscillator couplings. As a result, the eigenvalues of $W$ [cf. Eqs.~(\ref{eq:M_matrix}) and (\ref{eq:w_posv})] are given by 
\begin{equation}
\label{eq:w_negv}
w_j = \bar{\omega} +  \mathrm{Re}[\lambda_j] +\mathrm{i} (-g-\mathrm{Im}[\lambda_j]), \quad v<0.
\end{equation}
Here, the real part of the eigenvalues (as well as the corresponding  eigenstates and thus the measure $\mathcal{S}$) remains unchanged, 
while the imaginary part simply changes its sign. Thus, eigenstates that are decaying for $v>0$, are growing for $v<0$ and vice versa. 

\section{Results}
\label{sec:results}
In the following we first discuss the case of $N=2$ in Sec.~\ref{subsec:TwoOscillators}, which provides a clear picture of the basic mechanism underlying synchronization of linear oscillators interacting via dissipative couplings. Subsequently in Sec.~\ref{subsec:NOscillators}, we consider a ring of $N>2$ oscillators and show that also in this case synchronized motion may be achieved and follows similar arguments as before. 

\subsection{Two coupled oscillators ($N=2$)}
\label{subsec:TwoOscillators}

Without loss of generality, we may choose the scaled frequency differences of the two oscillators to be $\Delta_1=+\Delta$ and $\Delta_2=-\Delta$, such that matrix $M$ governing the dynamics [cf. Eq.~(\ref{eq:M_matrix})] is given by  
\begin{equation}
\label{eq:M_dimer}
M=
    \begin{pmatrix}
     \Delta & -\mathrm i \\
    -\mathrm i &  -\Delta
    \end{pmatrix}
\end{equation}
Here, we have chosen $v>0$. However, from the discussion in Sec.~\ref{subsec:OurModel}  we know that a negative value of $v$ simply results in a change of sign of the imaginary part of the eigenvalues.
The two eigenvalues and corresponding right eigenvectors of $M$ are given by
\begin{align}
\label{eq:lambda_dimer}
    \lambda_{\pm} &= \pm  \sqrt{\Delta^2-1}
    \\
    \vec{c}_\pm
    &=\frac{1}{\sqrt{1+|\Delta\pm\sqrt{\Delta^2-1}|^2}}
    \begin{pmatrix}
    \mathrm i(\Delta\pm\sqrt{\Delta^2-1}) \\
    1
    \end{pmatrix}
    \label{eq:eigvec_dimer}
\end{align}
If $|\Delta|<1$ ( $|\Delta|>1$) the eigenvalues $\lambda_{\pm}$ are both purely imaginary (real) and non-degenerate. In contrast, for $\Delta=\pm 1$ not only are the eigenstates degenerate but also the corresponding eigenvectors coalesce, i.e., these values of $\Delta$ correspond to exceptional points. 
The impact of exceptional points on synchronization goes beyond the scope of the present work and we will focus in the following on the cases $|\Delta|>1$ and $|\Delta|<1$.

\begin{figure}
    \centering
    \includegraphics[width=\columnwidth]{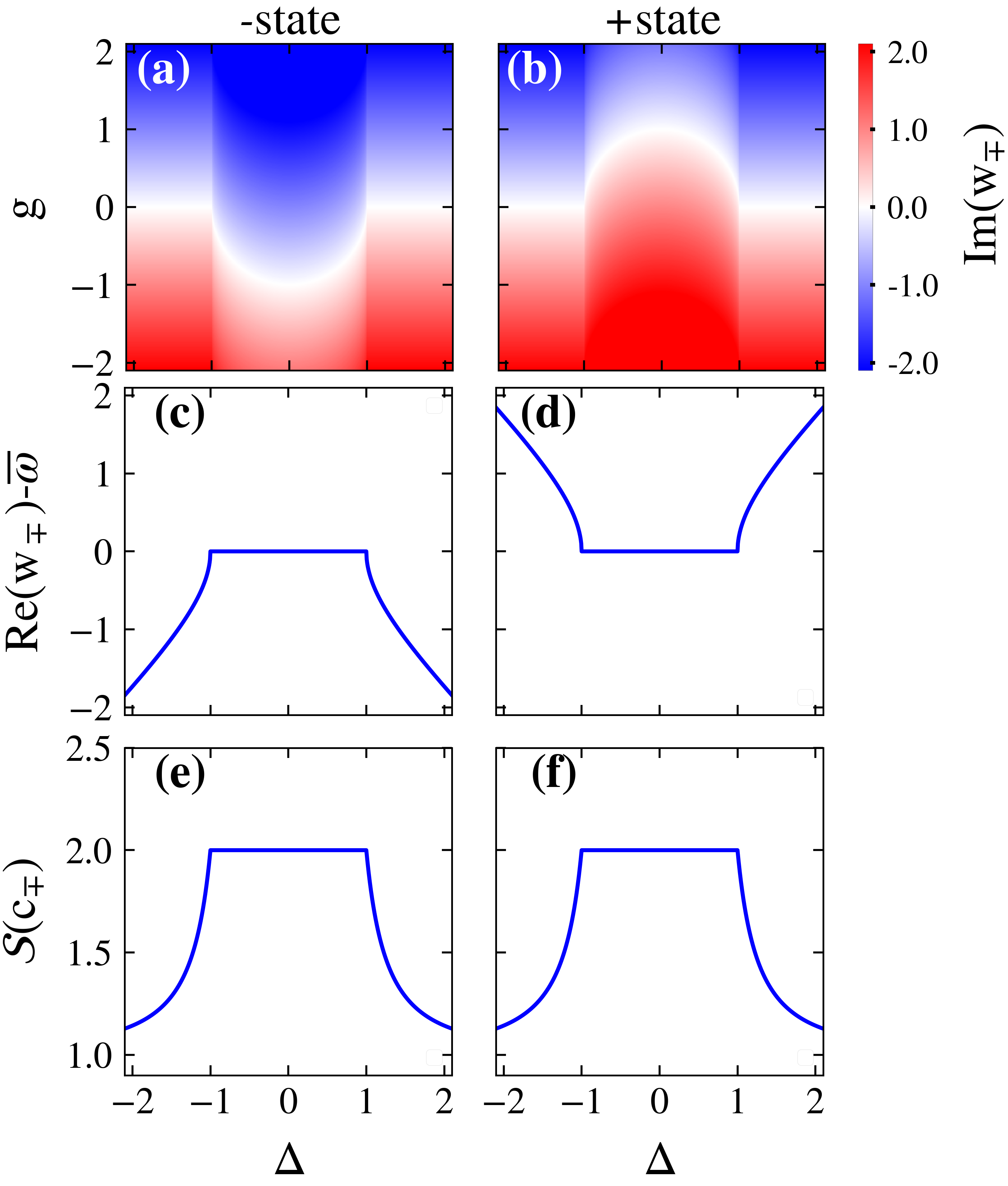}
   \caption{Top row: Density plots of the imaginary part $\mathrm{Im}(w_\pm)$ as a function of the frequency difference $\Delta $ and the dissipation strength $g$: (a) $w_-$ and (b) $w_+$. 
   Dissipation-free synchronization is found along the white line. Middle row: Corresponding real part (c) $Re(w_-)$ and (d) $Re(w_+)$ as a function of $\Delta$, which corresponds to the oscillation frequency of the respective eigenvector. Last row: Degree of synchronization $\mathcal{S}$ as function of $\Delta$ of the eigenvalue (e) $\vec c_-$ and (f) $\vec c_+$. The largest value is found for $|\Delta |<1$ corresponding to fully synchronized motion.}
    \label{fig:dimer}
\end{figure}

\paragraph{Overview:}
As discussed in Sec.~\ref{subsec:OurModel}, the eigenenergies $w_\pm = \bar{\omega} +  \mathrm{Re}[\lambda_\pm] + \mathrm{i} (-g+\mathrm{Im}[\lambda_\pm])$ describe the overall possibility of long lasting synchronized motion in terms of oscillation frequency and damping, while $\mathcal{S}$ quantifies the degree of synchronization. 
Let us start by considering the imaginary part of the eigenenergies $w_\pm$ given by $\mathrm{Im}[w_\pm]=  -g+ \mathrm{Im}[\lambda_\pm]$, which determines the (exponential) damping or growing. 
In Figs.~\ref{fig:dimer} (a) and (b) we show $\mathrm{Im}(w_-)$ and $\mathrm{Im}(w_+)$, respectively, as a function of the frequency difference $\Delta$ and the dissipation strength $g$. 
Note, that $\Delta$ as well as $g$ can take on positive and negative values. 
The red areas in Fig.~\ref{fig:dimer}(a) and (b) indicate positive values corresponding to amplitude growth whereas the blue areas indicate negative values and thus amplitude damping. 
The two regions are separated by a white region, where amplitudes neither increase nor decrease. 
We discuss this most relevant region for dissipation free synchronization in more detail below.

As expected from the discussion above, quite different behavior of $\mathrm{Im[w_\pm]}$ is observed depending on whether $|\Delta|>1$ or $|\Delta|<1$. 
Similarly, a pronounced difference is found in the behavior of the real part  $\mathrm{Re}[w_\pm]=  \bar\omega+ \mathrm{Re}[\lambda_\pm]$, which describes the oscillation frequency of the eigenmodes and is shown in Fig.~\ref{fig:dimer}(c) and (d). 
For $|\Delta|<1$ the frequency remains unchanged and both eigenstates oscillate with the mean frequency $\bar{\omega}$.  However, for $|\Delta|>1$ the frequency of the $-$~state [cf. Fig.~\ref{fig:dimer}(c)] is decreasing, while that of the $+$~state [cf. Fig.~\ref{fig:dimer}(d)] is increasing. Both follow the functional form of a square-root with opposite sign, cf. Eq.~(\ref{eq:lambda_dimer}). 
Lastly, in Fig.~\ref{fig:dimer}(e) and (f) we show the degree of synchronization  $\mathcal{S}$ as function of $\Delta$, which is given by [cf. Eq.~(\ref{eq:eigvec_dimer})]
\begin{equation}
    \label{eq:S_dimer}
    \mathcal{S}(\vec{c}_\pm,\Delta)=\left\{
    \begin{tabular}{lcl}
    2 & \quad ,\ & $|\Delta|<1$ \\
    $2 \frac{\Delta^2}{2\Delta^2-1}$
    & \quad ,\ & $|\Delta|>1$
    \end{tabular}\right..
\end{equation}
As expected, the maximum value lies within the range of $|\Delta|<1$ and rapidly decreases as $|\Delta|$ increases, indicating the absence of synchronization. After this broad overview we will in the following discuss in more detail the potential of synchronized motion in the system of $N=2$ oscillators, focusing on the three criteria (i)--(iii) formulated in Sec.~\ref{subsec:OurModel}.

\begin{figure}
    \centering
   \includegraphics[width=0.9\columnwidth]{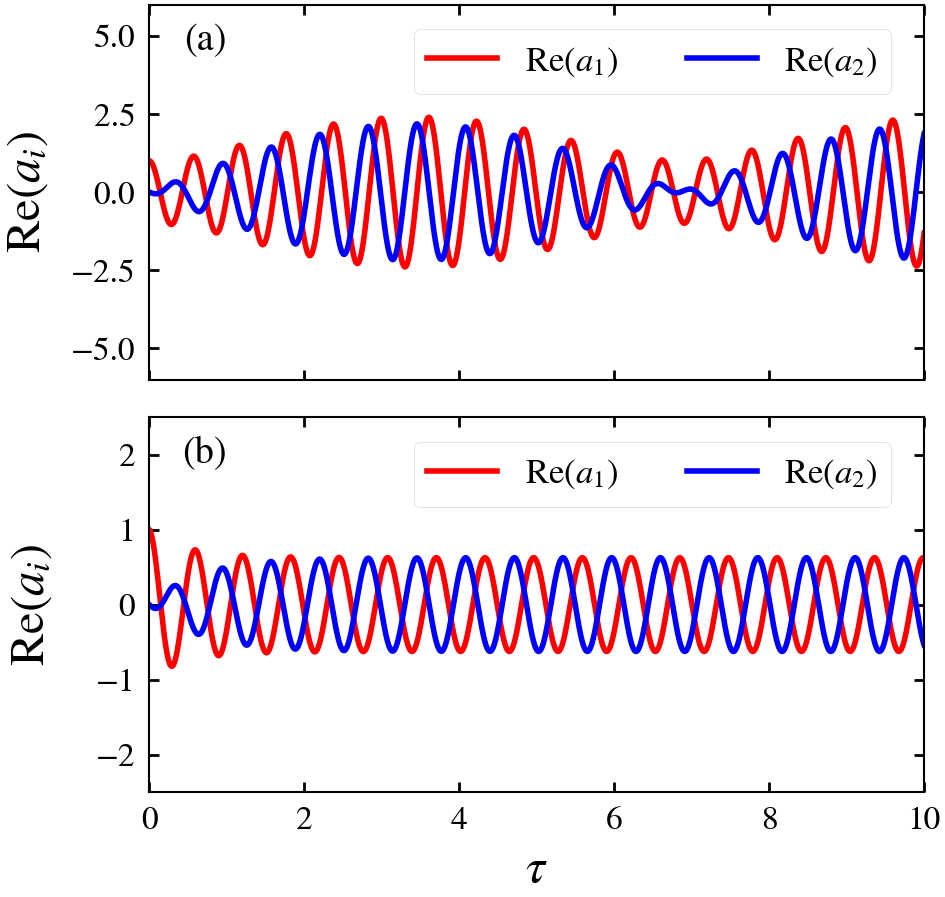}
   \caption{Examples of different dissipation free dynamics found for the case of $N=2$ oscillators. We plot the real amplitude $Re(a_{n}(\tau))$ of the first oscillator in red ($n=1$) and the second one in blue ($n=2$). (a) For $\Delta=1.1$ and $g=0$, the presence of two oscillation frequencies within the dissipation free subspace leads to beating. (b) For $\Delta = 0.6$ and $g=0.8$, only a single eigenstate with its respective oscillation frequency is dissipation free, while the other is damped leading to a periodic steady state of both oscillators, i.e., synchronization. Parameters: $\bar{\omega}=10$, $\vec{a}(0)=(1,0)^\intercal$. 
   These results are obtained by direct integration of the differential equation. It agrees perfectly with the results obtained via diagonalization.
   }
    \label{fig:dynamics}
\end{figure}

\paragraph{Detailed discussion of the regime $|\Delta|>1$:}
In this case, the eigenvalues $\lambda_\pm$ become purely real [cf. Eq.~(\ref{eq:lambda_dimer})], such that the eigenenergies take the simple form $w_\pm = (\bar{\omega}\pm \sqrt{\Delta^2-1}) - \mathrm{i} g$. Most importantly, the imaginary part is solely given by $-g$ for both states and is independent of $\Delta$, which can also be seen in Figs.~\ref{fig:dimer}(a) and (b). Thus, both eigenstates show the same dynamical response to dissipation, i.e., either both are dissipation free  ($g=0$) or the amplitudes decay/increase with the same rate given by $-g$.  
Although there exists a dissipation free subspace for $g=0$, and thus requirement (i) is fulfilled, requirement (ii) cannot be fulfilled simultaneously. The reasons is that both states have different oscillation frequencies $\bar{\omega}\pm \sqrt{\Delta^2-1}$ and none of them is decaying, resulting in a beating pattern. 
We show an example of such a  time evolution of the real amplitudes $\mathrm{Re}(a_n)$  governed by Eq.~(\ref{eq:dot_a_k_normalizedDelta}) in Fig.~\ref{fig:dynamics}(a)  for $\Delta = 1.1$ and $g=0$. 

\paragraph{Detailed discussion of the regime $|\Delta|<1$:}
After we have ruled out the possibility of synchronization [according to our conditions (i)--(iii)] in the previous regime, we now discuss the case of $|\Delta|<1$, where dissipation free synchronized  motion is indeed possible. For $|\Delta|<1$ the eigenvalues $\lambda_\pm$ are purely imaginary [cf. Eq.~(\ref{eq:lambda_dimer})] and dissipation free states are determined by $0=-g \pm \sqrt{|1-\Delta^2|}$, such that condition (i) may be fulfilled. 
In contrast to the previous case, we need to differentiate between the two states: Dissipation vanishes for the $+$~state if $g= g_{+} \equiv \sqrt{|1-\Delta^2|}$, and for the $-$~state if $g = g_{-}\equiv -\sqrt{|1-\Delta^2|}$. 
Each of these solutions describes a half circle with radius one, cf. Figs.~\ref{fig:dimer}(a) and (b).

We now examine whether condition (ii) is also fulfilled in this regime. When the $-$~state is dissipation free, the amplitude of the $+$~state is growing exponentially as $\mathrm{Im}[w_{+}(g_{-})]= -g_{-}+\sqrt{1-\Delta^2} = 2 \sqrt{1-\Delta^2}  >0$. This is also verified by Fig.~\ref{fig:dimer}: Along the white region in panel (a) within the regime $|\Delta|<1$, the area in panel (b) is red. In contrast, along the white region in panel (b), the area in panel (a) is blue, i.e. while the $+$~state is dissipation free, the $-$~state is damped. 
Specifically, $\mathrm{Im}[w_{-}(g_{+})]= -g_{+}-\sqrt{1-\Delta^2} = -2 \sqrt{1-\Delta^2}  <0$. 
Thus, synchronized motion for $|\Delta|<1$ is found whenever the condition $g= \sqrt{1-\Delta^2}$ is fulfilled. 
Moreover, this state has a degree of synchronization of $\mathcal{S}=2$ and is therefore fully synchronized for all $|\Delta|<1$. 

In Fig.~\ref{fig:dynamics}(b)  we show the dynamics for the parameters $\Delta=0.6$ and $g=0.8$ when starting in the initial state $\vec{a}(0)=(1,0)^\intercal$. 
As discussed previously, we expect to find synchronized motion for these parameters.
Indeed, after a short transient time of $\tau \gtrsim 2$ a stationary oscillatory motion emerges where both oscillators have the same amplitude. 
Note the phase shift between the two oscillators, which may be understood as follows: Considering the $+$~state $\vec{c}_{+}$  [cf. Eq.~(\ref{eq:eigvec_dimer})], the long time dynamics is given by $ \vec{a}_\mathrm{sync}(t) = \vec{c}_+ \exp[-\mathrm{i}\omega_+ t]$; cf. Eq.~(\ref{eq:time-evol_eigen}). Then, 
\begin{align}
    \mathrm{Re}[\vec{a}_\mathrm{sync}(t)] 
    =& \mathcal{N} \begin{pmatrix}
       \cos(\omega_+ t+ \phi) \\
       \cos(\omega_+ t)\\ 
    \end{pmatrix},
\end{align}
where the phase difference $\phi$ fulfills $\tan(\phi) = -\sqrt{1-\Delta^2}/\Delta$ and $\mathcal N = (1+|\Delta+\sqrt{\Delta^2-1}|^2)^{-1/2}$ is the normalization constant from Eq.~(\ref{eq:eigvec_dimer}).

\subsection{Many coupled oscillators on a ring}
\label{subsec:NOscillators}

In this section, we generalize our results from the previous Sec.~\ref{subsec:TwoOscillators} for the case of two coupled oscillators to large numbers of oscillators arranged on a ring. 
Also for the case of $N$ oscillators, the dynamics is  governed by Eqs.~(\ref{eq:dot_a_k_normalizedDelta})--(\ref{eq:w_posv}). In the following we will first discuss the case of equal frequencies of all oscillators. Afterwards, we discuss the more relevant case of frequency differences.

\subsubsection{Identical frequencies of all oscillators}
\label{subsubsec:IdenticalFrequencies}
To gain a basic understanding of the eigenstates and eigenvector structure we now consider the case when all frequencies are identical, i.e. $\Delta_n=\Delta$. Then, the eigenvalues and (right) eigenvectors of $W$ are given by 
\begin{align}
    \label{eq:wi_equal_delta}
    w_j &=(\bar\omega+\Delta) -i\Big(g \pm 2  \cos(\frac{2 \pi j}{N} )\Big), \quad v\gtrless 0, 
    \\
    \label{eq:ci_equal_delta}
    \vec{c}_j &= \frac{1}{\sqrt{N}} \sum_{n=1}^N e^{\mathrm{i} \frac{2 \pi}{N} j n}\vec{e}_n,
\end{align}
 where $\vec{e}_n$ is the $n$th unit-vector.
 As all eigenstates are independent of $\Delta$ or $g$.
One sees that most eigenstates are degenerate. For even $N$ only the eigenstates with $j=N$ and $j=N/2$ are not degenerate; for odd $N$ only the state with $j=N$  is not degenerate.
Moreover, the real part of the eigenenergies $w_j$, i.e. the oscillation frequencies, is simply shifted by $\Delta$ for all eigenstates.
However, the imaginary part of $w_j$, which dictates the dissipation and more importantly the possibility of dissipation free dynamics, requires a more careful analysis. 
 
\paragraph{Positive $v$:}
The imaginary part of the $j$th eigenvalue $\mathrm{Im}[w_j]=0$ if $g=g_j\equiv -2  \cos({2 \pi j}/{N} )$. Then, all other eigenvalues $w_{j'}$  with $j'\ne j$ have imaginary part given by
\begin{equation}
\label{eq:ImCos1}
\mathrm{Im}[w_{j'}(g_j)]= 2  \cos\left(\frac{2 \pi j}{N} \right) -2  \cos\left(\frac{2 \pi j'}{N} \right).
\end{equation}
Furthermore, we need to distinguish the two cases of odd and even $N$: For an \textit{odd} number of oscillators and $j\neq (N\pm1)/2$ there is always at least one $j'$ with $\mathrm{Im}[w_{j'}(g_j)]>0$, and thus condition (ii) is not fulfilled. On the other hand, if $j = (N\pm1)/2$ all other eigenstates are damped except for $j' = j\mp 1$. Yet, this state is also dissipation free and condition (ii) cannot be fulfilled. For \textit{even} $N$, however, there exists a non-degenerate eigenstate $j=N/2$ that fulfills (i) and (ii). Then, $g=2$ and 
$\vec{c}_\mathrm{syn}\equiv \vec{c}_{N/2}=\frac{1}{\sqrt{N}} (-1,1\dots,-1,1)^\intercal$, which corresponds to anti-phase synchronization between nearest neighbors with the same frequency $\bar\omega+\Delta$.

\paragraph{Negative $v$:}
In contrast to the previous case. the imaginary part of the $j$th eigenstate now is equal to zero if $g=g_j\equiv +2  \cos({2 \pi j}/{N} ) $ and thus Eq.~(\ref{eq:ImCos1}) becomes 
\begin{equation}
\mathrm{Im}[w_{j'}(g_j)]= -2  \cos\left(\frac{2 \pi j}{N} \right) +2  \cos\left(\frac{2 \pi j'}{N} \right)
\end{equation}
for all other eigenvalues $w_{j'}$ with $j'\ne j$. Here, only if $j=N$ are all other states damped and conditions (i) and (ii) fulfilled. 
The corresponding eigenstate is $\vec{c}_\mathrm{syn}\equiv \vec{c}_N=\frac{1}{\sqrt{N}} (1,\dots,1)^\intercal$, i.e., in-phase synchronization of all oscillators with frequency $\bar\omega+\Delta$.

\subsubsection{Oscillators with different frequencies}
In this section, we discuss the case of arbitrary frequency differences $\Delta_n$ for each oscillator on the ring. In this case, the matrix $M$ [cf. Eq.~(\ref{eq:M_matrix})] can no longer be diagonalized analytically. 
Therefore, we discuss the basic behavior along a few examples of $\Delta_n$ and solve the eigenvalue problem numerically. Yet, these examples demonstrate that dissipation free synchronized motion also exists in such a general setup.

A convenient way to investigate how the properties of synchronization are affected by changes of $\Delta_n$, is to parametrize the frequency difference according to 
\begin{equation}
    \Delta_n= s_n \Delta,
\end{equation}
and analyze the behavior of the eigenvalues and eigenvectors of $W$ as a function of $\Delta$ for a given (and fixed) set of $s_n$. Furthermore, we choose $v$ to be \textit{negative}, such that for $\Delta=0$ there exists a fully synchronized eigenstate if $g=2$ (see the discussion in Sec.~\ref{subsubsec:IdenticalFrequencies}b). Note that a negative value of $v$ implies $g_j=\mathrm{Im}[\lambda_j]$.

\begin{figure*}
    \centering
    \includegraphics[width=6.45cm]{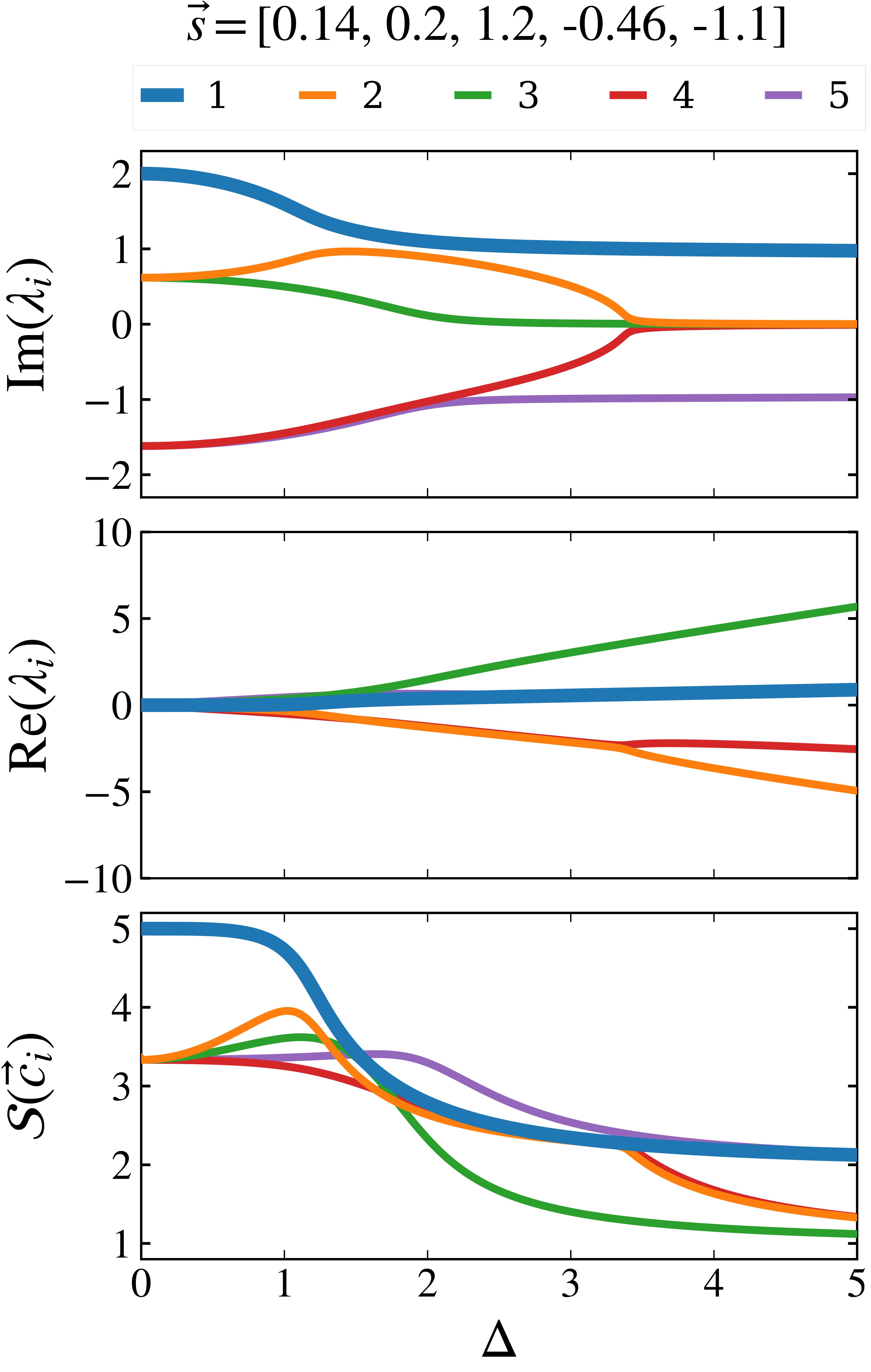}
    \includegraphics[width=5.5cm]{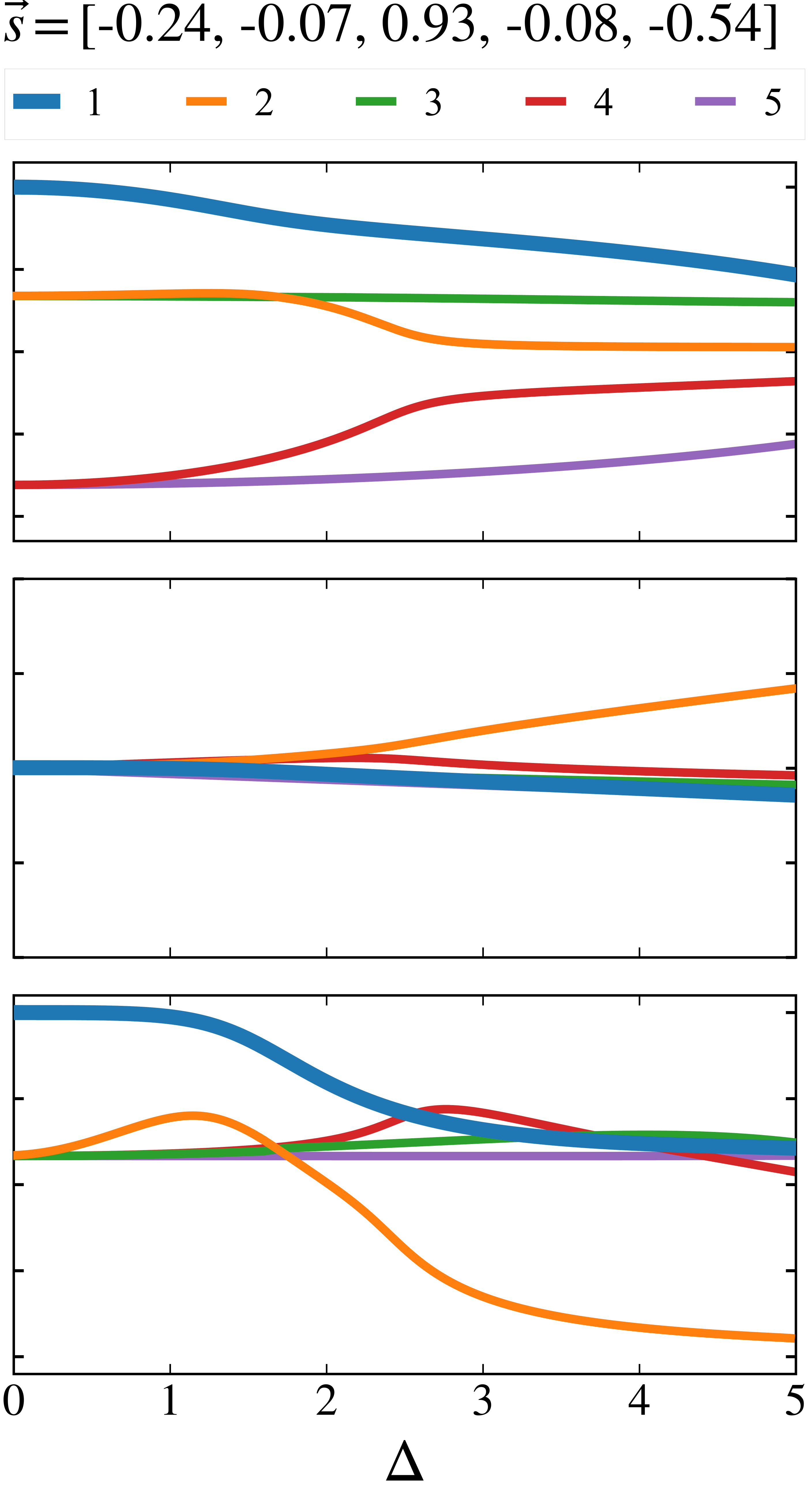}
    \includegraphics[width=5.5cm]{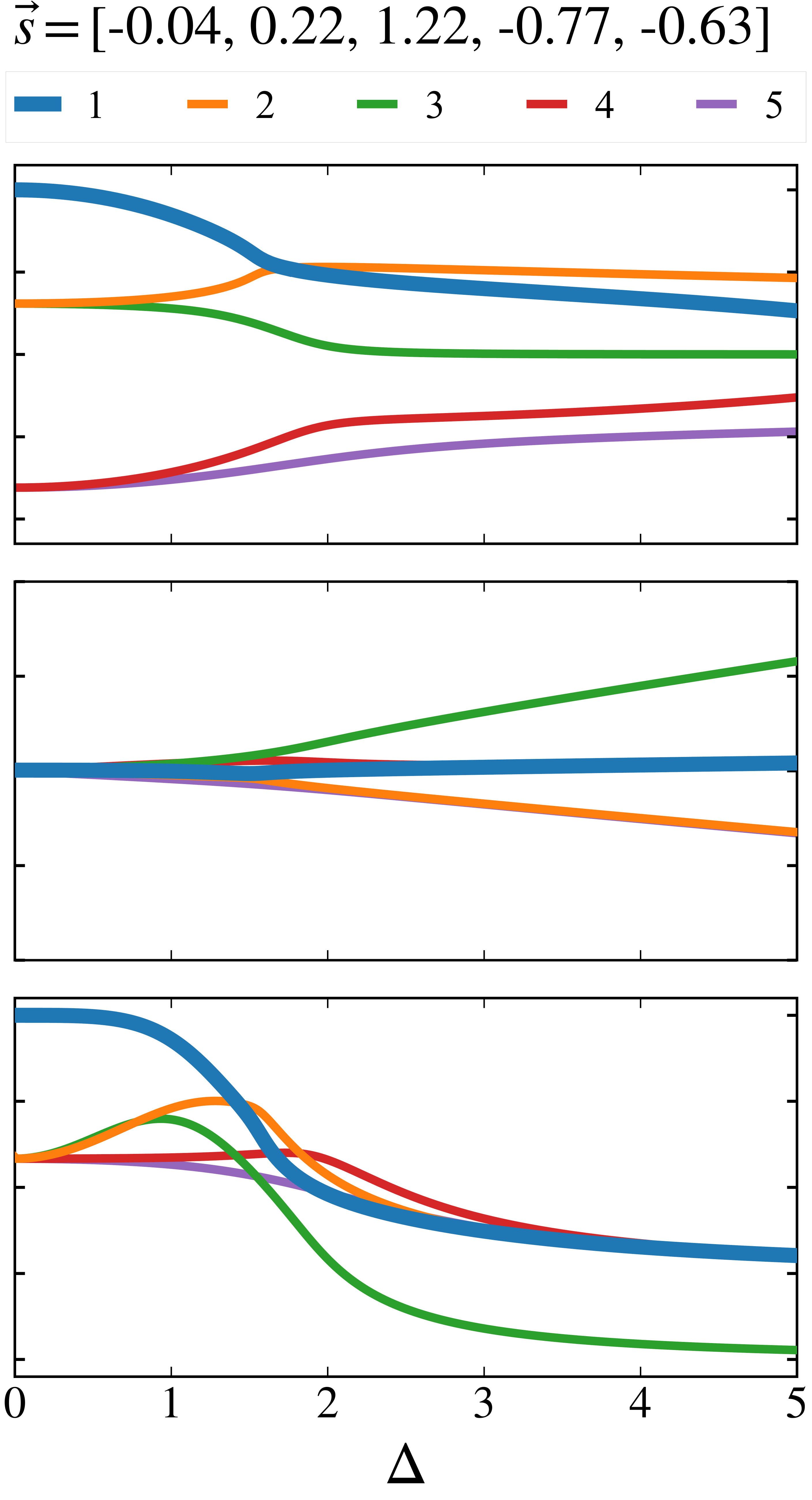}
   \caption{Examples of dissipation free and (fully) synchronized dynamics in a ring of $N=5$ oscillators with random frequency disorder. 
   The three different columns correspond to three different set of (scaled) frequency realizations $\vec s$. 
   The value of $v$ is taken to be negative. 
   In the top row we show the imaginary part $\mathrm{Im}[\lambda_j]$ of the eigenvalues $\lambda_j$ of the matrix $M$ as a function $\Delta$. 
   The middle row shows the corresponding real part $\mathrm{Re}[\lambda_j]$ and the bottom row the degree of synchronization $\mathcal S(\vec c_j)$ of the corresponding eigenstates $\vec c_j$. 
   For all three considered realizations, there exists an eigenstate (blue) with the maximum value of $\mathcal S$  (bottom row) for small values of $\Delta \lesssim 1$. 
   This eigenstate also has the largest imaginary part of its associated eigenvalue (top row), which allows the tuning $g$ in such a way that it becomes dissipation free while all other eigenstates are damped.}
    \label{fig:example_N=5}
\end{figure*}

In the following we consider as example the case of $N=5$ oscillators and show in Fig.~\ref{fig:example_N=5} the results of the numerical diagonalization of the matrix $M$ for three different realizations of $\vec s = (s_1, ..., s_5)$ (different columns). We choose the largest difference between neighboring values of $s_n$ to be equal to one, i.e.~$\max[s_n-s_{n+1}] = 1 $. 
Then, for $\Delta<1$ all frequency differences between neighboring oscillators are always smaller than the dissipative coupling between them (which has magnitude one). 

The case of $N=2$ in our network of oscillators allows us to represent the full parameter space as shown in Fig.~\ref{fig:dimer} and identify the dissipation free subspaces and synchronization within. 
However, for larger system sizes (as considered now)  a representation similar to Fig.~\ref{fig:dimer} becomes very space consuming. 
Yet, a dissipation free subspace is always necessary for synchronization, which corresponds to the white lines in Figs.~\ref{fig:dimer}(a) and (b). 
Thus, in order to determine whether conditions (i)--(iii) are fulfilled, it is sufficient to only search along the parameters for which each eigenstate becomes dissipation free. 
In particular, the relevant information of Fig.~\ref{fig:dimer}(a) and (b) may be conveniently combined to  contain only $g_\pm=\mathrm{Im}[\lambda_\pm]$ as function of $\Delta$. 
Accordingly, the top row of Fig.~\ref{fig:example_N=5} shows the imaginary part of all eigenvalues $\mathrm{Im}[\lambda_j]$ as function of the parameter $\Delta$ and the middle row shows the respective real parts $\mathrm{Re}[\lambda_j]$. 
Lastly, in the bottom row we plot the degree of synchronization $\mathcal S$ of each eigenvector also as function of $\Delta$. 
The eigenvalues of $M$ are sorted in descending order of their imaginary parts, i.e. $\mathrm{Im}[\lambda_{1}]>\mathrm{Im}[\lambda_{2}]>\dots>\mathrm{Im}[\lambda_{N}]$. 

In the following we discuss different regimes of $\Delta$ and its impact on the possibility of synchronized motion in accordance with conditions (i)--(iii). 
We focus on the eigenstate $\vec{c}_1$ with largest imaginary part $\mathrm{Im}[\lambda_{1}]$ (highlighted as thick blue lines in Fig.~\ref{fig:example_N=5}). 
The reason is that for $g=\mathrm{Im}[\lambda_{1}]$ the eigenstate $\vec{c}_1$ becomes dissipation free while all other eigenstates are simultaneously damped. 
In contrast, if we would choose $g$ such that another eigenstate $\vec{c}_{j\neq 1}$ would become dissipation free, there is at least one eigenstate that is exponentially growing. 
It is thus sufficient to only analyze the possibility of synchronization of $\vec{c}_1$ in the following. 

\paragraph{No frequency difference $(\Delta =0)$:} This means that there are no variations in the oscillator frequencies and the situation is exactly the same as discussed in Sec.~\ref{subsubsec:IdenticalFrequencies}b. Consequently, the eigenvalues of $W$ are given by Eq.~(\ref{eq:wi_equal_delta}). 
From the discussion in Sec.~\ref{subsubsec:IdenticalFrequencies}b, we know that if $g=2= \mathrm{Im}[\lambda_\mathrm{syn}]$ there exists a dissipation free synchronized state $\vec{c}_\mathrm{syn}\equiv\frac{1}{\sqrt{5}}(1,\dots,1)^\intercal$ with associated real eigenvalue $w_\mathrm{syn}=\bar{\omega}$, i.e. all oscillators are in phase and oscillate with frequency $\bar{\omega}$. This is exactly what we observe in Fig.~\ref{fig:example_N=5}:  the eigenvalue with largest imaginary part has imaginary part $\mathrm{Im}[\lambda_1]=2$ (blue  thick lines in the top row). Note that $\mathrm{Im}[\lambda_2]=\mathrm{Im}[\lambda_3]$ and $\mathrm{Im}[\lambda_4]=\mathrm{Im}[\lambda_5]$. Furthermore, $\mathrm{Re}[\lambda_j] = 0$ (middle row) which implies an oscillation frequency of $\bar{\omega}$.

\paragraph{Small frequency differences  $(0< \Delta < 1)$:}
In this regime, the disorder in the frequency differences between nearest neighboring oscillators always remains smaller than the coupling between them (which is $1$). 
We thus expect that the degree of synchronization also remains large [$\mathcal S(\vec{c}_1)\approx N$], i.e. the full delocalization of the eigenstate $\vec{c}_1$ persists. 
In the bottom row of Fig.~\ref{fig:example_N=5} we observe exactly this behavior of the thick blue line corresponding to $\vec{c}_1$: For small values of $\Delta$, $\mathcal S(\vec{c}_1)$ is maximal and slowly decreases as $\Delta$ approaches the value of $1$. 
Thus, the synchronized state remains close to be fully synchronized within this regime [condition (iii)]. 
Note, that the values for which $\mathcal S(\vec{c}_1)$ starts to decrease depends on the specific realization of disorder $\vec s$. 

The imaginary part of the corresponding eigenvalue (top row) continues to be the largest value of all eigenvalues (blue thick line), $\mathrm{Im}[\lambda_1]>\mathrm{Im}[\lambda_{j\neq 1}]$. Thus, for $g=\mathrm{Im}[\lambda_1]$ the eigenstate $\vec c_1$ becomes dissipation free while all other eigenstates are damped, i.e. conditions (i) and (ii) are fulfilled. 
As $\Delta$ increases, $\mathrm{Im}[\lambda_1]$ decreases resulting from the larger amount of frequency disorder. 
Simultaneously, the real part $\mathrm{Re}[\lambda_1]$ remains close to $0$ such that the oscillation frequency of the synchronized state $\vec{c}_1$ also continues to be close $\bar\omega$. 
Note, the value of $\mathrm{Re}[\lambda_1]$ only affects the oscillation frequency.

\paragraph {Large frequency differences $(\Delta \geq 1$):}
As $\Delta$ is increased further, the frequency difference exceeds the nearest neighbor interaction such that -- similar to (Anderson) localization in finite systems \cite{Moebi_arxiv} 
 -- the degree of synchronization $\mathcal{S}(\vec c_1)$ of the previously delocalized eigenstate $\vec c_1$ rapidly decreases as $\Delta$ increases; see blue thick lines in the bottom row of Fig.~\ref{fig:example_N=5}. 
Hence, only partial synchronization is possible in this regime and condition (iii) is not fulfilled. 

At the same time, the largest imaginary value $\mathrm{Im}[\lambda_1]$ continues to decrease as function of $\Delta$. 
Yet, close to $\Delta = 1$ it remains well separated from the second largest imaginary value  $\mathrm{Im}[\lambda_2]$ such that a suitable choice of $g$ still allows for dissipation free dynamics with a sinlge oscillation frequency. 
However, $\mathrm{Im}[\lambda_1]$ may  coalesce with $\mathrm{Im}[\lambda_2]$ for larger values of $\Delta$ depending on the specific realization of $\vec s$. 
An example of such a degeneracy is observed for $\Delta\approx 1.6$ in the top right panel of Fig.~\ref{fig:example_N=5}. 
As a result, both eigenstates would be dissipation free resulting in the beating pattern discussed previously in Sec.~\ref{subsec:TwoOscillators}. However, as mentioned above, only partial synchronization is possible in this regime anyways.

\paragraph {Very large frequency differences $(\Delta \gg 1$):}
In the regime of very large frequency differences, we expect that the degree of synchronization takes its minimum value $\mathcal{S}(\vec{c}_j)=1$ for all eigenstates $j$ since the scaling follows $\Delta \gg v$. 
This implies that the values $\Delta_n = \Delta s_n$ are much larger than the dissipative coupling strength $v$. Then, $M$ is approximately diagonal with eigenvectors $\vec{c}_j$ nearly localized. 
Note that in this limit there is no synchronized state. We have checked numerically that for $\Delta$ larger than the smallest difference between the $s_n$ the synchronization measure of all eigenstates approaches one, as expected (not shown here).

\begin{figure}
    \centering
    \includegraphics[width=7cm]{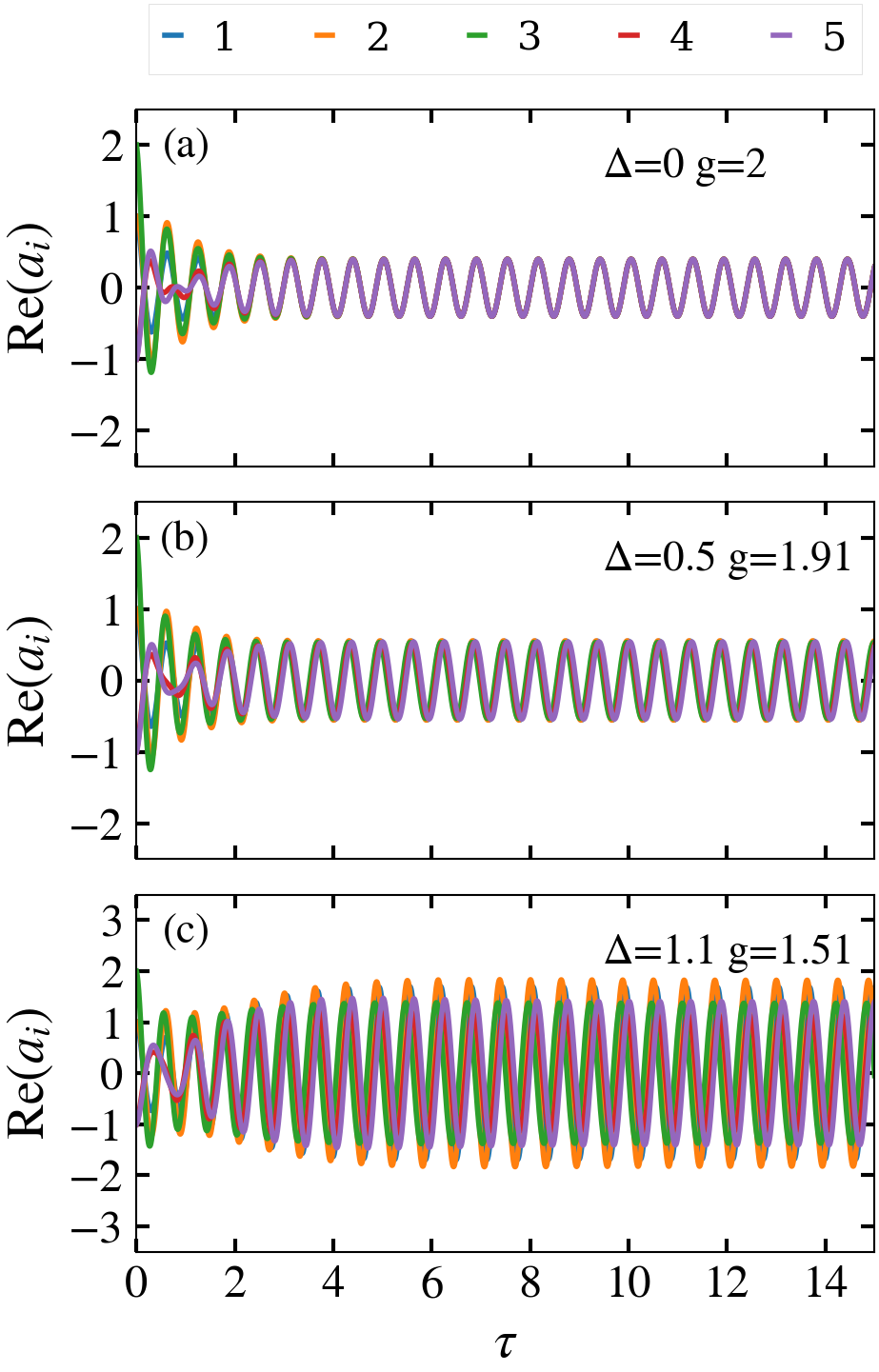}
   \caption{Dynamical behavior of $\mathrm{Re}(a_i(\tau))$ given by Eq.(\ref{eq:M_dimer}) for different values of the scaling factor $\Delta$. In all three cases the mean frequency of the oscillators is $\bar{\omega}=10$ and the disorder is the same of the first panel of Fig.~(\ref{fig:example_N=5}), namely $\vec{s}=(1.14,\ 0.20,\ 1.20,\ -0.46,\ -1.1)$ 
    The coupling strength $v$ is taken to be negative  and all frequencies are given in units of $|v|$. The initial condition is $\vec{a}_0=(1,1,2,-1,-1)$. Panels (a) and (b) show fully synchronized motion, while panel (c) is an example of partial synchronization. }
    \label{Fig.example_N=5_dyn}
\end{figure}

Lastly, to demonstrate that the dynamics of the system of oscillators is consistent with our discussion of the different regimes above (obtained from analyzing the eigenvectors and eigenfrequencies), we show in Fig.~\ref{Fig.example_N=5_dyn} examples of $\mathrm{Re}[a_n(\tau)]$ as a function of the scaled time $\tau$ for $\vec{s}=(1.14, 0.20, 1.20, -0.46, -1.1)$ (corresponding to the first column of Fig.~\ref{fig:example_N=5}) for three different values of $\Delta$.
In all cases, we choose the initial state $\vec{a}_0=(1,1,2,-1,-1)$.

Panel (a) corresponds to the case of vanishing frequency difference, i.e. $\Delta =0$. We choose the dissipation $g=2$ such that only the eigenstate with largest imaginary part is dissipation free. As expected after a short transient time  of $\tau \approx 2.5$ all oscillators are in-phase synchronized.

In panel (b), we increase the frequency difference to be $\Delta =0.5$. Hence, the synchronized state is dissipation free for $g=1.91$. Analogues to the previous case (a), all oscillators are synchronized after a transient time of $\tau \approx 2.5$, yet with a small phase shift. 
Importantly, all oscillators have the same amplitude consistent with the finding of Fig.~\ref{fig:example_N=5} that the degree of synchronization is maximal [$\mathcal S(\vec c_1)=5$ for this value of $\Delta$].

Contrarily, in panel (c) where $\Delta = 1.1$ (and $g= 1.51$ to match the condition of dissipation free dynamics) the amplitudes vary among the oscillators. This is in accordance with $\mathcal S(\vec c_1) <5$. However, still only a single oscillation frequency is present (after some transient time). This is an example of partial synchronization.

\section{Conclusions}
\label{sec:Conclusions}

In this work we have investigated the possibility of long-lived synchronized motion in networks of harmonic oscillators, which are subject to gain/loss and interact via nearest neighbor dissipative couplings. 
In this context, we refer to synchronization as the existence of a single eigenstate of the dynamical matrix, which is dissipation free.
Furthermore, if it attains the maximum value of the (inverse) participation ratio we refer to it as `fully synchronized'.
We find that in the case of only two coupled oscillators, synchronization may always be achieved by tuning the gain appropriately as long as the frequency difference between the two oscillators is smaller than their interaction strength. 

A similar behavior may be observed in larger networks, i.e.\ many oscillators arranged on a ring with nearest neighbor interactions, yet the possibility of synchronization then depends on the specifics of the system at hand: 
If all oscillators are identical, synchronized collective motion may be achieved for an even number of sites with repulsive dissipative couplings ($v$ positive) \emph{or} an odd number of sites with attractive dissipative interactions ($v$ negative). 
For small frequency differences compared to the coupling between the oscillators, this behavior remains, which we show specifically for the case of $N=5$, yet it should also hold for larger networks. 
However, as the number of coupled oscillators increases, it becomes increasingly difficult to achieve full synchronization and may only be observed for very small frequency differences.
For larger frequency differences, the (inverse) participation ratio decreases significantly such that only partial synchronization may be achieved. 
This is in accordance with Anderson localization, where on-site disorder results in localized eigenstates. 
However, as the dynamical matrix in this work is non-Hermitian, Anderson localization is not directly applicable. 
Here, future work is needed to study the interplay of synchronization and localization, in particular in the thermodynamic limit and arbitrary small frequency perturbations. 

Synchronization as discussed in this work is intimately related to the existence of dissipation free dynamics and thus isolated points/submanifolds in parameter space. 
Hence, they require a very precise tuning of gain and loss in order to obtain periodic steady states. 
This is however hard to achieve in any realistic experiment and the synchronized state will experience some gain or loss. 
We can relax the condition $\mathrm{Im}[\EValTot_j]=0$ by solely requiring $|\mathrm{Im}[\EValTot_j]|\ll |\mathrm{Re}[\EValTot_j]|$, which means that the change of amplitude of oscillation is small over many oscillations.
In addition, we then require $\mathrm{Im}[\EValTot_j]\ll \mathrm{Im}[\EValTot_\mathrm{sync}]$, which means that all other eigenstates decay much faster than the 'synchronized' one. 
In principle, one may relax the condition even further and demand that there exists only one state with $\mathrm{Im} [\EValTot_j]>0$, while all other states fulfill $\mathrm{Im}[\EValTot_i]\leq 0$. 
Then the synchronized state would grow while all other states are exponentially damped.

\begin{acknowledgements}
C.W.W.\ acknowledges support from the Max-Planck Gesellschaft via the MPI-PKS Next Step fellowship and is financially supported by the Deutsche Forschungsgemeinschaft (DFG, German Research Foundation) – Project No. 496502542 (WA 5170/1-1). 
A.E.\ acknowledges support from the DFG via a Heisenberg fellowship (Grant No EI 872/10-1).
\end{acknowledgements}

\bibliographystyle{apsrev4-1}

\end{document}